\documentclass[aps,pre,twocolumn,showpacs,superscriptaddress,amsmath,amssymb]{revtex4}

\usepackage{graphicx}
\usepackage{dcolumn}
\usepackage{bm}
\usepackage{color}
\usepackage{multirow}
\usepackage{textcomp}
\usepackage{umoline,txfonts}

\begin{document}

\title{Geometric properties of graph layouts optimized for greedy navigation}

\author{Sang Hoon Lee}
\affiliation{IceLab, Department of Physics,
Ume{\aa} University, 901 87 Ume{\aa}, Sweden}
\affiliation{Oxford Centre for Industrial and Applied Mathematics, Mathematical Institute, University of Oxford, Oxford OX1 3LB, United Kingdom}

\author{Petter Holme}
\affiliation{IceLab, Department of Physics,
Ume{\aa} University, 901 87 Ume{\aa}, Sweden}
\affiliation{Department of Energy Science, Sungkyunkwan University, Suwon 440--746, Korea}
\affiliation{Department of Sociology, Stockholm University, 106 91 Stockholm, Sweden}

\date{\today}

\begin{abstract}
The graph layouts used for complex network studies have been mainly been
developed to improve visualization. If we
interpret the layouts in metric spaces such as Euclidean ones, however, the embedded
spatial information can be a valuable cue for various purposes. In this work,
we focus on the navigational properties of spatial graphs. We use an recently user-centric
navigation protocol to explore spatial layouts of complex networks that are optimal for navigation.
These layouts are generated with a simple simulated annealing optimization technique. We compared these layouts to others targeted at
better visualization. We discuss the spatial statistical properties of the optimized layouts
for better navigability and its implication.
\end{abstract}



\maketitle

\section{introduction}
One of the quintessential properties of spatial structures is to support
human navigation, or give cities, buildings etc.\ their {\em navigability}.
Navigability in purely topological (not spatially embedded) graphs or networks
have been studied, especially during the last decade of complex network
research~\cite{ComplexNetwork}. Navigability in more relevant context of
spatial networks~\cite{SpatialNetwork}, however, is a more recent topic~\cite{NavigabilitySpatialNetwork}.
As a contribution to this topic, we have introduced greedy spatial navigation (GSN)
as a simple probe of the effects of human cognitive limitation to navigation in non-familiar environments~\cite{SHLee2011,SHLee2012}.
The essence of GSN is the quantification of navigability for {\em given} spatial structures.
In this work, we focus on the reverse problem of {\em giving} spatial
structures for better navigability.

In our previous work, we made the  observation that a spring-embedding
 layout (for better visualization) actually helps the greedy
navigators to find better routes compared to a random layout~\cite{SHLee2011}. The difference is occasionally very large,
depending on the network structure. This ``side effect'' comes
from the property that layouts for visualization usually put
topologically close vertex pairs geometrically close. In this paper,
we study the layouts optimized solely for the better greedy navigability.
In particular, we focus on geometric characteristics of optimized layouts
compared to the aforementioned visualization oriented layouts. The optimization is done by
simulated annealing (SA) process~\cite{SA}. This will be further discussed in the next section.

\section{layout optimization procedure}
We begin the optimization procedure by assigning a random position
for each vertex independently, inside  square of unit length. The object
function to minimize is the average steps greedy navigators
take for all the vertex pairs as source--target ($s$--$t$) ones (denoted as $d_g$
 in Ref.~\cite{SHLee2012}).
We use the SA~\cite{SA} technique that repeatedly
applies the heating and quenching processes. At each time step,
the position of a randomly chosen vertex, whose coordinates are $(x_0,y_0)$, is relocated to
$(x_0 + \Delta x, y_0 + \Delta y)$ where $\Delta x$ and $\Delta y$
are randomly drawn from the interval $[-l,l]$ ($l \gg 1$) uniformly. In the heating (quenching) process, such a trial movement is accepted if $d_g$ is
decreased, while it is accepted with the probability $p_\textrm{high}$ ($p_\textrm{low}$)
otherwise, respectively. This is similar to successful  heuristic methods in combinatorial
optimization~\cite{Ardelius2006}.
The trial movement in the heating process is repeated
for $T_H$ times in the unit of Monte Carlo (MC) steps, where one step is defined as
$N$ (the number of vertices) trial movement. On the other hand, in the quenching process,
the trial movement is repeated until consecutive $T_L$ times
(again, in the unit of MC steps) of rejection occurs. Overall, these consecutive heating
and quenching processes as a single session are repeated $T_{HL}$ times in total.
Finally, during this procedure, the layout $L_\textrm{min}$ corresponding to the minimum $d_g$ value
up to present is recorded and constantly updated if a new minimum $d_g$ occurs.
With the method, after the SA procedure ends, we obtain the approximated optimal
layout for GSN up to the moment.

\section{results}

\begin{figure}
\includegraphics[width=\columnwidth]{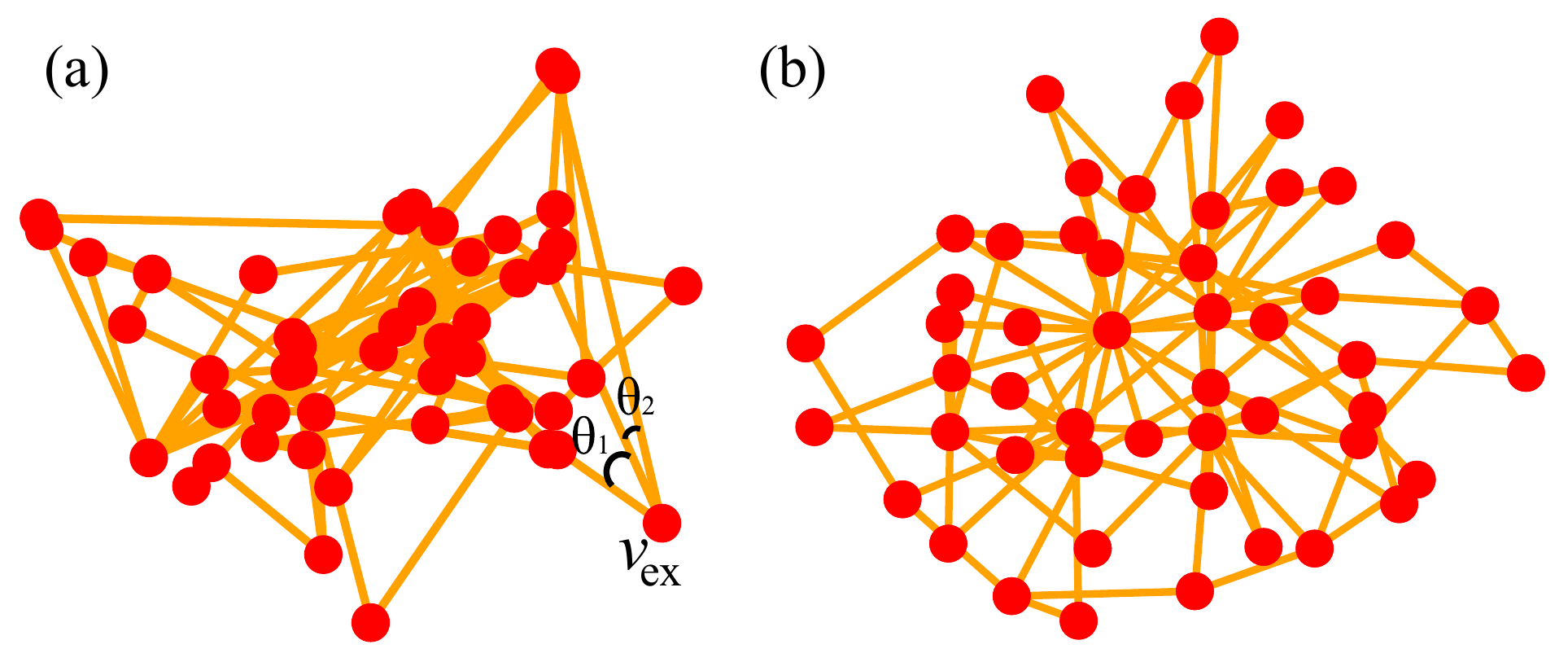}
\caption{(color online) Examples of the optimal (a)
and KK layout (b) of the BA model. The GSN pathway is $3.85$ ($4.79$) for the optimal
(KK) layout, respectively.
}
\label{BA_example}
\end{figure}

\begin{figure}
\includegraphics[width=\columnwidth]{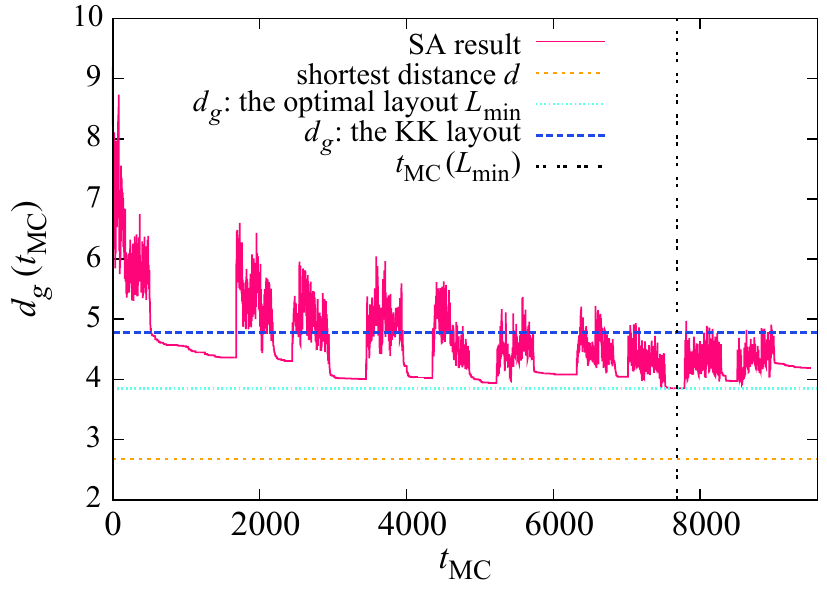}
\caption{(color online) A typical time series of $d_g$ in the unit of MC steps $t_\textrm{MC}$,
in case of the BA model used in Fig.~\ref{BA_example}, along with the real shortest
path length $d$, $d_g$ for the optimal layout $L_\textrm{min}$ [Fig.~\ref{BA_example}(a)]
and the KK layout [Fig.~\ref{BA_example}(b)],
and the moment of $L_\textrm{min}$ denoted as the vertical line.
}
\label{BA_time_series_example}
\end{figure}

To start our exploration, we set $l = 10$, $p_\textrm{high} = 0.2$, $p_\textrm{low} = 0$
(completely frozen phase), $T_H = 500$, $T_L = 100$, and $T_{HL} = 10$.
Due to the computational complexity we present the results using the graphs with rather small sizes:
$N = 50$, in case of Barab{\'a}si--Albert (BA)~\cite{BA}, Holme--Kim (HK)~\cite{HK},
Watts--Strogatz (WS)~\cite{WS} model graphs, where the average degree is
set as $\bar{k} = 2$ for BA and HK and $\bar{k} = 4$ for WS graphs (triangle formation probability
for HK model is $1$ and the rewiring probability for WS model is $0.1$).
As an example of real-world graph, we analyze the social network of the oft-studied
Zachary karate club with $N = 34$~\cite{KarateClub}. As representative examples
of completely regular structures, the two-dimensional square lattice (2D square) with the open boundary condition
with $N = 7 \times 7 = 49$ and the one-dimensional ring (1D ring) with $N = 50$ and $k = 2$
(connected only with the nearest neighbors) are analyzed as well.
A typical example of $L_\textrm{min}$ and time series of $d_g$
in case of the BA model is illustrated in Figs.~\ref{BA_example}(a) and \ref{BA_time_series_example}.

\begin{figure}
\includegraphics[width=0.7\columnwidth]{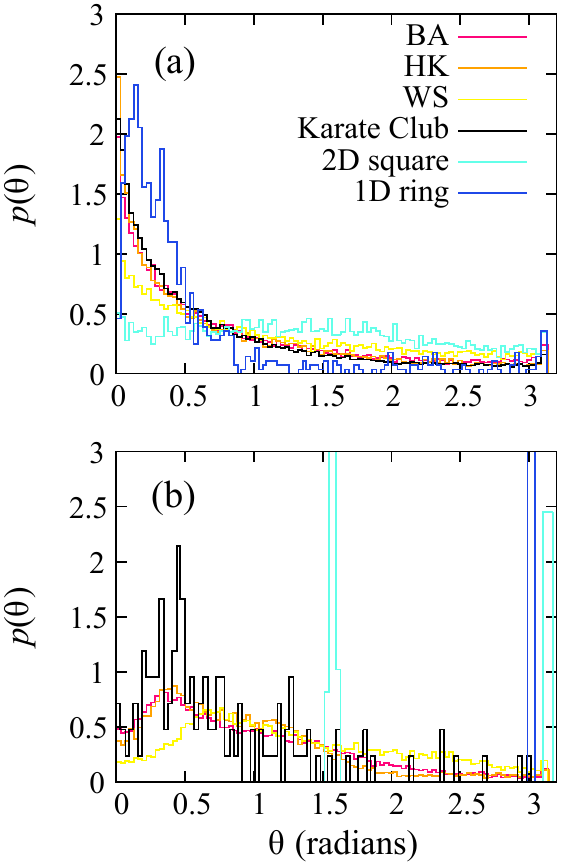}
\caption{(color online) Distribution of angles in the optimized (a) and KK (b) layout.
At least $39$ graph ensembles are used to average for all the cases.
}
\label{AngleDist_opt_KK}
\end{figure}

\begin{figure}
\includegraphics[width=0.7\columnwidth]{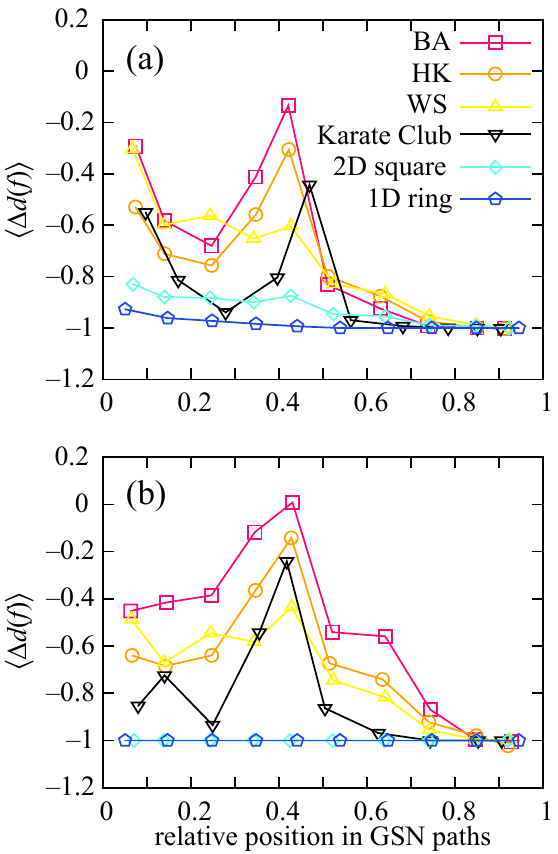}
\caption{(color online) Average step decreased along the GSN paths in terms
of the relative position $f$ along the pathways, in the optimized (a) and KK (b) layout.
At least $39$ graph ensembles are used to average for all the cases,
and the horizontal axis is equipartitioned into $10$ bins.
}
\label{dx_avg}
\end{figure}

In our further analysis we first investigate the angle (defined as the undivided angles between edge pairs attached to each vertex---in Fig.~\ref{BA_example} this is means we include $\theta_1$ and $\theta_2$ but not $\theta_1 + \theta_2$)
distributions of optimal layouts for GSN,
in comparison to the spring-embedding Kamada--Kawai (KK) layout for the purpose of visualization~\cite{KK},
as shown in Fig.~\ref{AngleDist_opt_KK}. An example of $L_\textrm{min}$ and time series of $d_g$
in case of the KK layout is shown in
Figs.~\ref{BA_example}(b) and \ref{BA_time_series_example}. Even if it is demonstrated that the
KK layout is helpful for GSN compared to the random navigation~\cite{SHLee2011},
the optimal layout shows significantly different angular profiles to the
ones in KK layout. In particular, much sharper angles dominate the former case,
in contrast to the latter case composed of a certain range of characteristic angles
for better visualization. Among the optimized layouts, the graphs with hubs
(vertices with large numbers of neighbors) such as the BA, HK, and Karate Club graphs, are obviously
dominated by more sharper angles attached to the hubs than the WS graph case.

For a deeper investigation of GSN pathways, we analyze the average
distance (number of edges in the shortest path to the target) decreased
along the GSN pathway. We  denote this quantity as $\langle \Delta d(f) \rangle$ and
present in Fig.~\ref{dx_avg}, where the relative position in a pathway
is denoted as the fraction $f = n/d_g$ ($n = 1, \cdots , d_g - 1$)
for the $n$'th intermediate vertex.
As seen, for all the graphs,
$| \langle \Delta d(f) \rangle |$ is larger for the KK layout
than for the optimal layout in the early stage of the GSN pathways.
However, the optimal pathway shows larger $| \langle \Delta d(f) \rangle |$
soon afterwards, which leads to better performance (shorter GSN pathways).
Another aspect is the peak of $\langle \Delta d(f) \rangle$
for both the KK and optimal layouts, indicating the characteristic
intermediate peak (point of inefficiency) in the GSN pathways.

In contrast to the random graph models (BA, HK, and WS) and Karate Club graph,
in case of the regular lattices (2D square and 1D ring) the KK layout provides
the exactly optimal layout for GSN, where the topological square lattice is mapped to
a geometrical square lattice with almost right angles. This can be seen from the peaks of Fig.~\ref{AngleDist_opt_KK}(b).
The topological 1D ring network is laid out as a circle (with almost straight angles as shown in the peak
in Fig.~\ref{AngleDist_opt_KK}(b)], and by  $\langle \Delta d(f) \rangle = -1$ regardless of $f$ in Fig.~\ref{dx_avg}(b)).
The SA algorithm, however, shows a suboptimal
performance for such regular structures as shown in Figs.~\ref{AngleDist_opt_KK}(a) and \ref{dx_avg}(a),
in contrast to the KK layout very good at detecting such regular structures.

\begin{figure}
\includegraphics[width=0.8\columnwidth]{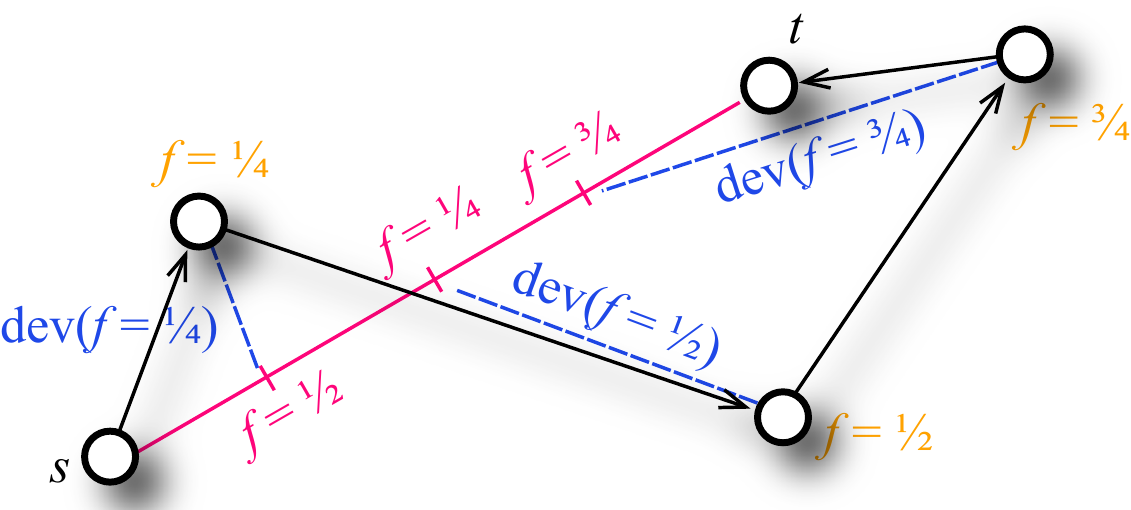}
\caption{(color online) An illustration of the definition of
deviation from the straight line for a $s$--$t$ pair. Each intermediate
vertex's location in the GSN pathway of length $4$
(arrows) is compared to the corresponding intermediate points
(green dots) equally spaced in the straight line (red line).
The deviation for each intermediate point is defined as the Euclidean
distance between the two points (blue dashed lines), in the unit of
the straight line.
}
\label{dev_schematic}
\end{figure}

\begin{figure*}
\includegraphics[width=0.9\textwidth]{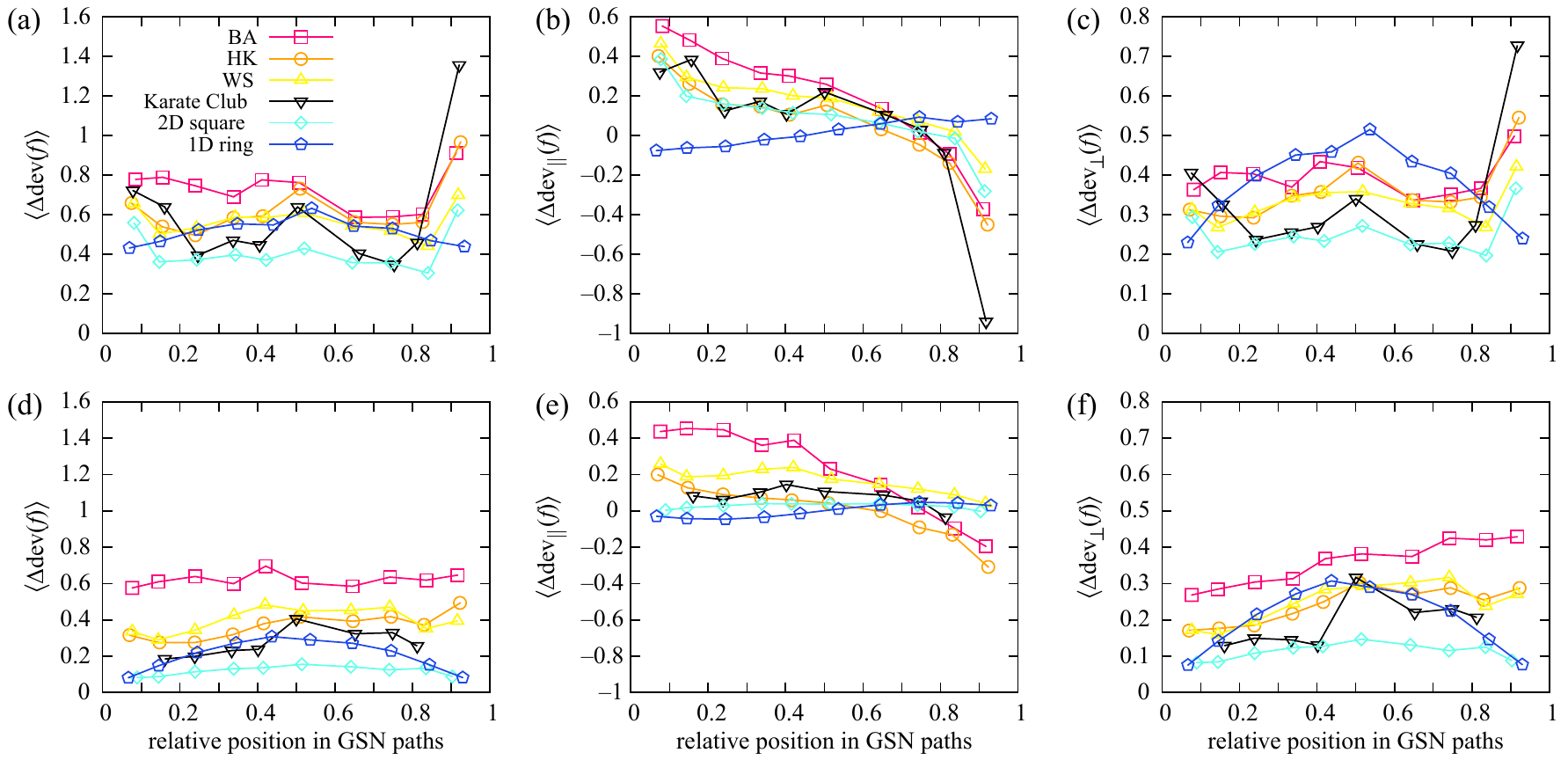}
\caption{(color online) Average deviation from the straight line
connecting $s$--$t$ pairs, for the relative position $x$
in the GSN pathways and straight lines.
Upper (lower) panels correspond to the optimized (KK) layouts,
respectively. $\langle \textrm{dev}(f) \rangle$ [(a) and (d)] is decomposed into $\langle \textrm{dev}_{\parallel}(f) \rangle$
[(b) and (e)] and $\langle \textrm{dev}_{\perp}(f) \rangle$ [(c) and (f)] with respect to
the straight line.
At least $39$ graph ensembles are used to average for all the cases.
The horizontal axis is binned into $10$ equidistant intervals.
}
\label{dev_from_lines}
\end{figure*}

Another way to explore the GSN pathways is to measure their straightness,
by comparing the pathways to the straight line connecting $s$--$t$ pairs.
Similar to the analysis of $\langle \Delta d(f) \rangle$, we quantify the
local straightness as a function of $f$, as illustrated in Fig.~\ref{dev_schematic}.
The deviation from the straight line connecting a $s$--$t$ pair, represented as a vector $\bold{v}_{st}$ from $s$ to $t$, at $f$
is denoted as $\textrm{dev}(f)$---the Euclidean distance between the intermediate
vertex at $f$ and the point at the same fraction $f$ on the straight line.
Note that all the distance measures for each $s$--$t$ pair are scaled
with respect to the length of the straight line ($|\bold{v}_{st}| = 1$), to facilitate the comparison
between all the different paths and layouts. Furthermore, each vector ${\bold{dev}}(f)$ (whose direction is from the
point on the straight line to the vertex on the GSN pathway) for the
distance $\textrm{dev}(f)$ is decomposed into the component parallel to
the straight line $\textrm{dev}_{\parallel}(f)$ and the one perpendicular to
it $\textrm{dev}_{\perp}(f)$. Due to the intrinsic directionality from the source
to the target, $\textrm{dev}_{\parallel}(f)$ can be positive or negative
depending on the angle between $\bold{v}_{st}$ and ${\bold{dev}}(f)$,
and we take the positive value of $\textrm{dev}_{\perp}(f)$ for convenience.

For the optimized and KK layouts applied to the graphs we have used,
values of average deviation from the straight lines as a function of $f$
are presented in Fig.~\ref{dev_from_lines}---$\langle \textrm{dev}(f) \rangle$, $\langle \textrm{dev}_{\parallel}(f) \rangle$,
and $\langle \textrm{dev}_{\perp}(f) \rangle$ separately. From the results,
we observe that in most cases the GSN pathways with KK layouts show much less
deviation from the straight lines than with the optimized layouts.
At a first glance, it may seem counterintuitive, because the GSN routes
are shorter in the optimized except for 2D square and 1D ring cases.
It is, however, reasonable if one recalls the fact that the path length
is defined as the hopping distance (number of edges along a pathway),
so that optimized layouts can have much longer edges compared to the
KK ones [compare Figs.~\ref{BA_example}(a) and (b)]. Such longer edges
amplify the entire length scale of optimized layouts compared to KK ones.
Therefore, in case of hopping-distance-based optimized layouts, the absolute
values of deviation themselves are not quite comparable to different types
of layouts. However, for a single type of layout, we can compare the effects
of different graph topologies. For instance, the (purely topological) BA model shows
the largest deviation in most cases, related to the amount of embedded geometric
information~\cite{SHLee2011}. The circular trajectories
of GSN pathways in the 1D ring case are clearly reflected in the concave curves
of $\textrm{dev}(f)$ and $\textrm{dev}_{\perp}(f)$ and the slightly negative (positive)
$\textrm{dev}_{\parallel}(f)$ for small (large) $f$, respectively,
for both optimized and KK layouts.
It would be interesting if we adopt the alternative definition of path length,
for instance, the sum of Euclidean distance along the path~\cite{SHLee2012a}
and compare the results.

\section{summary and discussions}
We have investigated the properties of optimized spatial graph layouts for
GSN generated by a simple SA process. From the observed geometrical measures,
it is shown that, in general, the optimized layouts are characterized by sharp angles
and point of inefficiency in the middle of the GSN processes. These
properties are qualitatively different from the layouts for better visualization,
namely the KK ones, showing the dominating intermediate angles within
characteristics ranges. A closer inspection of the navigational routes also
reveals that the inefficiency of KK layouts soon follows its initially better
performance compared to the optimized ones.
In other words, our simulation shows that it is
possible to generate GSN-friendly layouts other than the visualization-friendly
layouts by just taking the simple SA optimization process.

Our simple SA optimization process, however, also shows its limitation by
yielding the suboptimal results for completely regular structures such
as the 1D ring and 2D square lattice, where the KK layouts happen to coincide with
the exact solution of optimization for GSN. The SA optimization certainly
gets close to the exact optimal layout, but not perfectly, at least
in our simulation setting.
Besides such possible suboptimal performances, the observed properties
of optimized layouts for better navigation can, we believe, give valuable hints
for constructing various spatial structures in practice, e.g.,
the urban planning and architecture~\cite{Hillier1984}.
Adopting more sophisticated optimization processes and studies on
diverse graph structures would be a natural candidate for the
future work, for even better understanding of the
tripartite relationship of topology-geometry-navigability.

\begin{acknowledgments}
This research is supported by the Swedish Research Council and the WCU program through NRF Korea funded by MEST R31--2008--10029 (PH).

\end{acknowledgments}


\end{document}